\documentclass [prb,reprint,notitlepage,superscriptaddress,floatfix] {revtex4-1}
\usepackage{amsmath}
\usepackage{graphicx}
\usepackage{lmodern}
\usepackage{amsmath}

\usepackage{color}
\usepackage{amssymb}
\newcommand{\beq} {\begin{equation}}
\newcommand{\eeq} {\end{equation}}
\newcommand{\bea} {\begin{eqnarray}}
\newcommand{\eea} {\end{eqnarray}}
\newcommand{\be} {\begin{equation}}
\newcommand{\ee} {\end{equation}}
\renewcommand{\(}{\left(}
\renewcommand{\)}{\right)}
\renewcommand{\[}{\left[}
\renewcommand{\]}{\right]}

\DeclareMathOperator{\sgn}{sgn}

\newcommand{\su} {{\mathrm{SU(2)}}}

\begin{document}

\title{Superconducting and charge-density-wave orders in the spin-fermion model: a comparative analysis}
\author{Yuxuan Wang}
\affiliation{Department of Physics, University of Wisconsin, Madison, WI 53706, USA}
\author{Andrey Chubukov}
\affiliation{William I. Fine Theoretical Physics Institute,
and School of Physics and Astronomy,
University of Minnesota, Minneapolis, MN 55455, USA}
\date{\today}

\begin{abstract}
We present a comparative analysis of superconducting and charge-density-wave orders in the spin-fluctuation scenario for the cuprates.
 That  spin-fluctuation exchange gives rise to d-wave superconductivity  is well known. Several groups recently argued that the same
  spin-mediated interaction may also account for charge-density-wave order with momenta $(Q,0)$ or $(0,Q)$, detected in underdoped cuprates.
  This has been questioned on the basis that charge-density-wave channel mixes fermions from both nested and anti-nested  regions on the Fermi surface,
  and fermions in the anti-nested region do not have a natural tendency to form a bound state, even if the interaction is attractive.
   We show that anti-nesting is not an obstacle for charge order, but to see this one needs to go beyond the conventional Eliashberg approximation.
   We show that in the prefect nesting/antinesting case, when the velocities of hot fermions are either parallel or antiparallel,  the onset temperatures in superconducting and charge-density-wave channels are of comparable strength for any magnetic correlation length $\xi$.
   The superconducting $T_{\rm sc}$ is larger than $T_{\rm cdw}$, but only numerically.  When the velocities of hot fermions are not strictly parallel/antiparallel,
   $T_{\rm cdw}$ progressively decreases as $\xi$ decreases and vanishes
   at some critical $\xi$.
   \end{abstract}
\maketitle

\section{Introduction}

The experimental discovery of static charge-density-wave (CDW) order in underdoped cuprates~\cite{ybco,ybco_2,ybco_1,X-ray, X-ray_1,davis_1,mark_last} has re-ignited theoretical studies of the mechanism of
 CDW instability and its interplay with d-wave superconductivity. The CDW order accounts for a number of properties of the pseudogap phase
   in the cuprates and the understanding of the mechanism of CDW instability is an essential step towards the understanding of the pseudogap.

   Several groups~\cite{ms,ms_njp,efetov,pepin_pdw,pdw} analyzed charge order with diagonal momentum $(Q, \pm Q)$, which we refer to as CDW-diag, and compared it with with superconductivity 
   within the spin-fluctuation (hot spot) scenario. 
   It has been argued that the hot spot model has an approximate $\su$ particle-hole symmetry, which makes the onset temperatures for CDW-diag and superconducting order almost equal
    for large magnetic correlation length.  It has been proposed \cite{efetov, pepin_pdw}  that the pseudogap may be due to the fact that  over a wide range of $T$
    the system cannot determine between near-degenerate  superconducting and CDW-diag orders.  
    
  It turned out, however, that the CDW order in the cuprates has  momenta ${\bf Q}=(Q,0)$ or $(0,Q)$ along $X$ or $Y$ directions in the momentum space rather than along the diagonals.  We will follow Ref.~\onlinecite{mike} and refer to this order as CDW-x. The experimental value of ${\bf Q}$ for CDW-x order in the cuprates is close to the distance between hot spots~\cite{X-ray}.
      Fermions in the hot regions are the ones which mostly participate in magnetically-mediated interaction, and
     the closeness of the experimental ${\bf Q}$ to the distance between hot spots fueled speculations that the same magnetic fluctuations which favor charge order with diagonal momenta $(Q,Q)$  may also be responsible for CDW-x instability~\cite{laplaca,charge,debanjan,flstar,pepin_pdw}.

   The magnetic scenario for CDW-x order is appealing by two reasons. First, the CDW-x order between hot fermions develops together with pair-density-wave (PDW) order~\cite{pepin_pdw, pdw, kivelson, patrick,agterberg,fradkin_new,greco},
 and the combination of the two explains specific features of ARPES data~\cite{charge,patrick,coex, flstar}. Second, a Ginzburg-Landau (GL) analysis shows
 ~\cite{charge,tsvelik,pdw,debanjan,nie} that
 CDW-x order breaks not only $U(1)$ translational symmetry, as is expected for any incommensurate charge order with a complex order parameter, but also $C_4$ lattice rotational symmetry and time-reversal symmetry. The breaking of $C_4$ is the consequence of the fact that  CDW/PDW order develops in the form of stripes, and the breaking of time-reversal symmetry is the consequence of the fact that CDW-x order parameters between the pairs of hot fermions with center of mass momentum ${\bf k}$ and ${-\bf k}$
  develop with relative phase $\pm \pi/2$. These two order parameters transform into each other under time reversal, and by selecting one of these states, the system spontaneously breaks time-reversal symmetry. Both $C_4$ symmetry breaking and time-reversal symmetry breaking  have been observed in the experiments~\cite{davis_1,kerr,armitage,sidis,greven,louis_last}.

 The GL analysis assumes that CDW-x order does develop at a finite $T_{\rm cdw}$, and that this temperature is comparable to $T_{\rm sc}$ for d-wave superconducting (SC) instability, which also develops between hot fermions due to magnetically-mediated interaction~\cite{scalapino}. The mean-field value of $T_{\rm sc}$ is expected to be larger than
  $T_{\rm cdw}$ simply because superconductivity necessarily involves pairs of fermions with opposite directions of Fermi-velocities (nesting), while CDW-x instability involves
   pairs of fermions whose Fermi velocities are generally at some finite angle with respect to each other.  Still, if mean-field values of
   $T_{\rm sc}$ and $T_{\rm cdw}$ are comparable, the effects
    beyond mean-field (e.g., the pre-emptive breaking of discrete symmetries for CDW-x order) may lift $T_{\rm cdw}$ above $T_{\rm sc}$.  This calls for a detailed comparative analysis of the onset temperatures of SC and CDW-x orders at the mean-field level, by which we mean ladder approximations for the corresponding vertices (see Fig.~2).

   In a recent paper~\cite{charge}, the two of us compared the onset temperatures of SC and CDW-x instabilities by analyzing the structure of the  diagrammatic series
      for SC and CDW-x vertices in the quantum-critical regime, when the magnetic correlation length $\xi$ is infinite and the fermionic self-energy $\Sigma (\omega)$ has a non-Fermi liquid form $\sqrt{\omega}$ and exceeds the bare $\omega$ term at small frequencies~\cite{acs}.  We considered the generic case of some finite angle between Fermi velocities of hot fermions separated by $(Q,0)$ or $(0,Q)$ and found that kernels of the ``gap" equations are logarithmical in both SC and CDW-x channels, and the prefactor for the logarithm for CDW-x channel is only numerically smaller than the one for SC channel. We cut the logarithms by temperature and found that
      $T_{\rm sc}$ and $T_{\rm cdw}$ are of comparable strength at infinite $\xi$. At a finite $\xi$, the logarithm in $T$ in the CDW channel is cut
        already at $T=0$, and, as a result,
       $T_{\rm cdw}$ decreases and vanishes at some finite $\xi$. The SC instability, on the other hand, is not critically affected by decreasing $\xi$ and, in the absence of impurity scattering, survives at any $\xi$.

       Although this analysis is plausible, the cutting of the logarithm by temperature is not a
      rigorously justified procedure in the quantum-critical regime
        because  the logarithm in a particular cross-section of the ladder series is cut by
        $T$ (or $\xi^{-1}$) only if we set the  frequencies of external fermions $\pm\omega_m$ to zero.
         At a finite $\omega_m$
         the logarithms in the SC and CDW-x channels are already cut by $\omega_m$ even at $T=0$ and $\xi=\infty$, and one has to go beyond the leading
         logarithmical approximation to rigorously analyze the emergence of SC and CDW-x (see below).

           In this communication we present the alternative
        analysis of SC and CDW-x instabilities, which does not rely on a comparison of logarithms in the perturbation theory.
        Specifically, we show that at infinite $\xi$ the full linearized equation for CDW-x order parameter,
         from which one extracts the temperature of CDW-x instability, differs from the corresponding equation for the SC order parameter
         only by the strength of the effective coupling. We use recent non-perturbative results~\cite{max_new, future} for SC instability in the quantum critical regime which show, among other things, that the instability develops at any value of the coupling, and relate $T_{\rm sc}$ and $T_{\rm cdw}$. To see this, we focus on the seemingly ``worst case scenario" for CDW-x instability, when the Fermi surface is horizontal or vertical in hot regions and
        CDW order involves a half of fermions in the nested region and a half in the anti-nested region (see Fig.~1).
         We note in passing that an almost nested/antinested
        Fermi surface agrees well with ARPES data for Bi$_2$Sr$_2$CaCu$_2$O$_{8+x}$ (the ratio of antiparallel/parallel velocities for hot spots 1 and 2 in Fig.~1 is $13.6$, same for parallel/antiparallel velocities at hot spots 3 and 4, see Ref.~\onlinecite{mike}).
          For a homogeneous d-wave SC instability, nesting/anti-nesting is not an issue because the two fermions in the particle-particle channel necessarily have momenta $k$ and $-k$ and, hence, the same $\epsilon_k$.
        For CDW-x instability, the kernel of the ``gap" equation in the nested region is  same as in the SC channel, i.e., the product of the two Green's function $G$'s has (without self-energy) the same form $1/(\omega^2 + \epsilon^2_k)$ as the product of the two $G's$ in the SC channel.
        On the other hand, the product of the
        two $G$'s for CDW-x in the anti-nested region gives  $1/(i\omega_m -\epsilon_k)^2$, and this combination by itself
        does not lead to Cooper logarithm after integration over $\epsilon_k$ and $\omega_m$ because it contains a double pole  as a function of either
        $\omega_m$ or $\epsilon_k$.  However, the true kernel in the anti-nested region contains the product of two $G$'s in the combination with the bosonic propagator, and the latter also depends on $\epsilon_k$ and $\omega_m$ and contains poles (as a function of $\epsilon_k$) and  branch cuts (as a function of $\omega_m$) in both half-planes of complex $\epsilon_k$ and $\omega_m$. It then becomes an issue whether the contribution from the poles/branch cuts in the bosonic propagator
         yields the result comparable in magnitude to the one in the nested region.

        To analyze this issue, we do the same trick as was  recently used in the analysis of the optical conductivity in the cuprates~\cite{optical} and
         re-express the set of two coupled equations  for CDW-x order parameters in the nested and anti-nested regions
         as the single  equation for the \mbox{CDW-x} order parameter in the nested region with the effective interaction from a second-order composite process involving fermions in the anti-nested region (see Fig.~2).  The effective composite interaction $\chi_{\rm com}$ involves two $G$'s [the ones whose product gives $1/(i\omega -\epsilon_k)^2$] and two spin-fluctuation propagators $\chi (k, \omega)$.  We evaluate the product by integrating over ${\bf k}$ and $\omega_m$
         and compare $\chi_{\rm com}$  with the interaction in the SC channel, which is a single spin-fluctuation propagator.
         We show that the effective interaction is comparable to the original $\chi$, both in the Fermi liquid regime at moderate $\xi$ and in the quantum-critical regime at large $\xi$. That the two interactions are comparable by magnitude may seem strange because $\chi_{\rm com}$ would vanish if we approximated spin-fluctuation propagators by their values between the particles on the Fermi surface, as it is done in the Eliashberg approximation.  However, this approximation is rigorously justified for electron-phonon interaction, for which corrections to Eliashberg approximation are small in the ratio of phonon velocity to Fermi velocity,
          while for electronic pairing mechanism it is  justified only in the artificial limit of large number of fermionic flavors $N$ (Refs.~\onlinecite{acs,acf,acn,ms,senthil_1}), which we do not impose here. For the physical case of $N=1$, there is no parameter which would allow one to neglect the dependence on $\epsilon_k$ in the spin-fluctuation propagators. Keeping these dependencies, we find that the integral which determines $\chi_{\rm com}$ is non-zero
          due to the poles in the two bosonic propagators, considered as functions of $\epsilon_k$, and, moreover, its magnitude is  comparable to the
           original spin-fluctuation propagator $\chi$.

            We analyze both Fermi-liquid and quantum-critical regimes and show explicitly that $T_{\rm cdw}$ for CDW-x order is non-zero
            and differs from SC $T_{\rm sc}$ only by a numerical factor which increases as  magnetic correlation length $\xi$  gets larger.
               We then analyze the effect of deviation from perfect nesting/antinesting and show that $T_{\rm cdw}$ gets progressively
                 reduced upon decreasing $\xi$ and eventually vanishes above some critical correlation length.

 \section{The model}

 \begin{figure}
\includegraphics[width=0.8\columnwidth]{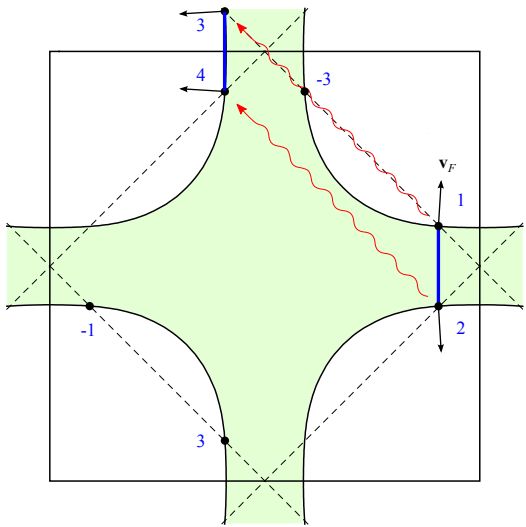}
\caption{The Brillouin zone, the magnetic Brillouin zone, and FS using the same dispersion as in Ref.\ \onlinecite{mike}. hot spots 1,2 and 3,4 are defined as points on the Fermi surface that intersects with the magnetic Brillouin zone. The red wavy lines represents the interaction mediated by spin fluctuations of momentum $(\pi,\pi)$. Hot spots -1, -2, -3, -4 (not shown) have momenta opposite to those of hot spots 1,2,3,4, respectively.}
\end{figure}

We consider the same model as in earlier studies~\cite{acs,efetov,acf,acn,ms,haslinger,berg,peter}: fermions in hot regions,
    interacting by exchanging  Landau-overdamped magnetic fluctuations peaked at $\bf K =(\pi,\pi)$.
   For the bulk of the paper we assume that the Fermi velocities of hot fermions separated by $(Q,0)$ or $(0,Q)$ are either parallel or antiparallel.
We approximate the dispersion of hot fermions by $\epsilon_k = v_y k_y$ and $\epsilon_k = -v_y k_y$  for fermions in  regions 1 and 2 in Fig.~1, and
    by $\epsilon_k = -v_x k_x$  for fermions in regions 3 and 4 in Fig.~1.

 Magnetically-mediated interaction $\Gamma_{\alpha \gamma, \beta \delta} ({\bf q}, \Omega_m)$ is proportional to dynamical spin susceptibility $\chi ({\bf k}, \Omega_m)$:
\beq
\Gamma_{\alpha\gamma,\beta\delta}({\bf k},\Omega_m)= {\bar g} \vec \sigma_{\alpha\beta}\cdot \vec\sigma_{\gamma\delta}~ \chi ({\bf k} - {\bf K}, \Omega)
\eeq
The static susceptibility $\chi ({\bf k} - {\bf K}, 0)$ comes from non-critical high-energy fermions (i.e., fermions with energies of order bandwidth),
and its momentum dependence is fully analytic and can approximated by a conventional Ornstein-Zernike form $1/({\bf k}^2 + \xi^{-2})$.  The dynamical part (the Landau damping) comes from low-energy fermions and must be computed together with fermionic self-energy $\Sigma (k, \omega_m)$. To simplify presentation, we follow earlier works~\cite{acf,acs,haslinger} and neglect the momentum dependence of the self-energy, i.e., approximate $\Sigma (k, \omega_m)$ by $\Sigma (\omega_m)$.
 We comment on this approximation below.
  For $\Sigma (k, \omega) \approx \Sigma (\omega)$,  self-consistent evaluation of the polarization operator and the fermionic self-energy then yields~\cite{acs, ms}
\beq
\chi ({\bf k} - {\bf K}, \Omega) \equiv  \chi (k, \Omega)  = \frac{1}{{\bf k}^2+\gamma|\Omega|+\xi^{-2}},
\label{a_1}
\eeq
with the Landau damping coefficient $\gamma = 4 {\bar g}/(\pi v^2_F)$ and
\begin{align}
\Sigma(k,\omega_m) &\approx \Sigma (k_h, \omega_m) \nonumber\\
&= \sgn(\omega_m) \sqrt{\omega_0} \[\sqrt{|\omega_m|+ \omega_{\rm sf} }- \sqrt{\omega_{\rm sf}}\]
\end{align}
where $k_h$ is a momentum of a hot fermion, $\omega_0 = {9 {\bar g}}/{(16\pi)}$, and $\omega_{\rm sf} = \xi^{-2}/\gamma = (\pi/4) (v_F \xi^{-1})^2/{\bar g}$.
This self-energy interpolates between Fermi liquid form at the smallest frequencies and quantum-critical, non-Fermi liquid form at larger frequencies:
\begin{align}
\tilde\Sigma(\omega_m)=\begin{cases}
\frac{3g\xi}{4\pi v_F} \omega_m\equiv \lambda\omega_m,&\text{for }|\omega_m|\ll\omega_{\rm sf}\\
\sgn(\omega_m)\sqrt{|\omega_m| \omega_0},&\text{for }|\omega_m| \gg \omega_{\rm sf}
\end{cases},
\label{cases}
\end{align}
where we have defined a dimensionless parameter
\beq
\lambda=\frac{3\bar g\xi}{4\pi v_F}.
\label{a_2}
\eeq
 The Green's functions of hot fermions are [${\tilde \Sigma} (\omega_m) = \Sigma (\omega_m) + \omega_m$]:
    \beq
    G_1(k, \omega) = \frac{1}{i {\tilde \Sigma} (\omega_m) - v_F k_y },   G_2(k, \omega) = \frac{1}{i  {\tilde \Sigma} (\omega_m) + v_F k_y},
    \label{y_1}
    \eeq
     for fermions in  regions 1 and 2 in Fig.~1, and
    \beq
    G_3(k, \omega) = \frac{1}{i  {\tilde \Sigma} (\omega_m) + v_F k_x},   G_4(k, \omega) = \frac{1}{i  {\tilde \Sigma} (\omega_m) + v_F k_x},
    \label{y_2}
    \eeq
 for fermions in  regions 3 and 4 in Fig.~1.

We now comment on the approximation $\Sigma (k, \omega_m) \approx \Sigma (\omega_m)$.
 First, despite that  $\Sigma (k_h,\omega_m)$ is parametrically larger than $\Sigma (k - k_h,0)$ at $\lambda \geq 1$
  (the difference is a power of $\lambda$, see~Ref.~\onlinecite{acs}), $\Sigma (k - k_h,0)$ is not small compared to $v_F |k-k_h|$, i.e., the
  renormalization of the  Fermi velocity of a hot fermion is not weak.  At the same time, the renormalization of the Fermi velocity
  does not generate a distinction between SC and CDW-x channels and  from this perspective is irrelevant for our consideration.  Second, the coupling $\lambda$ does actually depend on the location of ${\bf k}_F$ along the Fermi surface and diverges
   at $\xi = \infty$ only at a hot spot~\cite{acs,ms,acn}.  This momentum dependence does affect the values of $T_{\rm sc}$ (Refs.\ \onlinecite{wang,acn}) and of $T_{\rm cdw}$  but in non-crucial way, i.e., it affects the numbers but does not impose qualitative changes.

\section{SC and CDW-{x} instabilities}

 \begin{figure}
\includegraphics[width=.86\columnwidth]{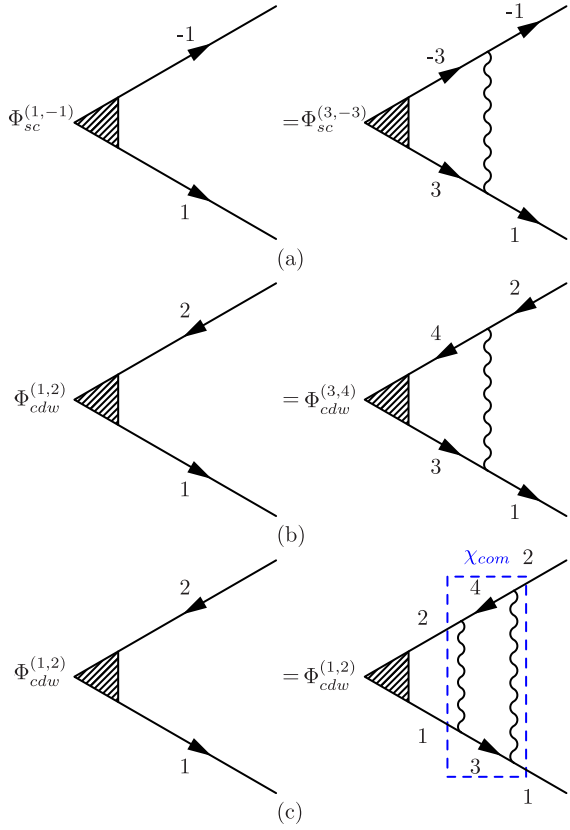}
\caption{The superconducting vertex $\Phi_{\rm sc}$ [Panel (a)] and the CDW-x vertex $\Phi_{\rm cdw}$ [Panels (b) and (c)] in the ladder approximation. Panel (b) shows the relation of CDW condensate formed by fermions at hot spots 1 and 2 with the one formed by fermions at hot spots 3 and 4. In turn, CDW condensate formed by fermions at  hot spots 3 and 4 is related in a similar way to that at hot spots 1 and 2, which we do not show here.
In Panel (c) we integrated out fermions near hot spots 3 and 4 and obtained the self-consistent equation for the CDW condensate formed by fermions at hot spots 1 and 2.}
\label{vs}
\end{figure}

To compare SC and CDW-x instabilities, we consider linearized equations for SC and CDW-x order parameters
 and compare at what temperature these equations have non-trivial solutions.
The equation for SC order parameter involves conventional ladder series in the particle-particle channel [see Fig.~2(a)]. Each cross-section contains two fermionic Green's functions with equal $\epsilon_k$ and opposite frequencies and one spin-fluctuation propagator.  The equation for CDW-x order parameter
is a $2\times 2$ set of coupled ladder equations
 for  order parameters in hot regions 1-2 and 3-4.  Each cross-section still contains the product of two $G$'s and one $\chi$, however in region 1-2 the
 two fermions  have equal frequencies and opposite $\epsilon_k$, while in region 3-4 they have equal frequencies and equal $\epsilon_k$.
 The  product of the two $G$'s in the region 1-2 is the same as the product of the two $G$'s in the SC cross-section, up to overall minus sign. The sign change is compensated by the summation over spin indices within the cross-section~\cite{ms}: in the spin-singlet SC channel, the spin structure of the particle-particle vertex is $i \sigma^y$, and $i\sigma^y_{\alpha \gamma} \vec\sigma_{\alpha\beta}\cdot \vec\sigma_{\gamma\delta} = (-3)i\sigma^y_{\beta \delta}$, while in CDW-x channel the spin structure of the particle hole-vertex is a $\delta-$function, and   $\delta_{\alpha \gamma} \vec\sigma_{\alpha\beta}\cdot \vec\sigma_{\delta\gamma} = (+3) \delta_{\beta \delta}$.  As a result, the kernel in region 1-2 in the CDW-x channel is equivalent to that in the SC channel.  Given this equivalence, it is convenient to
  re-arrange the ladder series in the $2\times 2$ set for \mbox{CDW-x} order parameter and represent then as a single ladder series involving fermions in the region 1-2 with the effective interaction $\chi_{\rm eff}$ coming from  second-order composite process in which fermions from the region 1-2 scatter into the region 3-4 and then scatter back into the region 1-2.
  This $\chi_{\rm eff}$ is the convolution of the two $G$'s in the region 3-4 and two bosonic $\chi$ with internal momenta/frequencies in the region 3-4 and external momenta/frequencies in the region 1-2, see Fig.~2(c).
 This composite $\chi_{\rm eff}$ has to be compared with the original $\chi$ at relevant frequencies and momenta.

 It is instructive to consider separately the case of moderate $\xi$, when the pairing involves fermions with energies below $\omega_{\rm sf}$, and the case of large enough $\xi$, when relevant fermionic energies exceed $\omega_{\rm sf}$.

  \subsection{SC and CDW-x instabilities in the Fermi-liquid regime}

 At energies below $\omega_{\rm sf}$, fermionic self-energy has a Fermi liquid form $\Sigma (\omega) = \lambda \omega$ and the momentum and frequency integral of the product of the two fermionic $G$'s in either SC channel or in CDW-x channel in the  region 1-2 yields a
  conventional Cooper logarithm $\log ({\omega_{\rm sf}/T})$. The
   logarithm comes from
   the smallest fermionic $\omega$ and $\epsilon_k$, hence, to logarithmic accuracy, $\omega$ and $\epsilon_k$ can be set to zero in the interactions.
       The linearized equations for superconducting and CDW condensates, $\Phi_{\rm sc} (k_x)$ and $\Phi_{\rm cdw} (k_x)$  are (in both cases $k_x$ is along the FS)
\begin{widetext}
   \bea
   & \Phi_{\rm sc} (k_x) = \frac{\lambda}{1 + \lambda} \log({\omega_{\rm sf}/T}) \frac{1}{\pi \xi}\int dk'_x \Phi_{\rm sc} (k'_x) \chi (k_x-k'_x,0) \nonumber\\
   &\Phi_{\rm cdw} (k_x) = \frac{\lambda}{1 + \lambda} \log({\omega_{\rm sf}/T}) \frac{1}{\pi \xi} \int dk'_x \Phi_{\rm cdw}  (k'_x) \chi_{\rm com} (k_x,k'_x,0),
   \label{ch_a}
   \eea
   \end{widetext}
   where $\lambda$ is given by (\ref{a_2}) and the  shift by ${\bf K}$  is absorbed into the definition of $\chi$ in
   Eq. (\ref{a_1}).
   To get the overall sign in the r.h.s. of the equation for $ \Phi_{\rm sc} (k_x)$ positive, we additionally assumed that SC order parameter has d-wave symmetry, in which case $\Phi_{\rm sc} ({\bf k}) =
   -\Phi_{\rm sc} ({\bf k} + {\bf K})$.
    [For CDW-x channel, $\Phi_{\rm cdw} (k)$ has opposite sign in the regions 1-2 and 3-4, but not equal magnitude. This  implies that the form-factor for
     CDW-x order is an admixture of $s-$wave and d-wave components (an admixture of a true CDW order and a bond charge order), and that d-wave component is larger~\cite{laplaca,charge,debanjan}.]

        The composite $\chi_{\rm com} (k_x,k'_x,0)$ is the convolution of two dynamical spin susceptibilities
   and two Green functions of fermions  with parallel velocities. Because we already have $\log {\omega_{\rm sf}/T}$ in the prefactor in (\ref{ch_a}), we
    can evaluate $\chi_{\rm com} (k_x,k'_x,0)$ at $T=0$, by replacing the summation over Matsubara frequencies by integration.
\bea
&&\chi_{\rm com} (k_x,k'_x,0) = -\frac{3\bar g}{8\pi^3}\int \frac{d\Omega_m dp_x dp_y}{(i\Omega_m (1 + \lambda) -v_Fp_x)^2} \nonumber \\
 &&\times \chi (k_x-p_x,p_y, \Omega) \chi (k'_x-p_x, p_y, \Omega)
 \label{y_3}
\eea
 where momenta $p_x, p_y$ and frequency $\Omega_m$ are for fermions in region 3-4.

Because the integrand contains double pole, a regularization is required. It is provided by either taking the external frequency for the vertex $\Phi_{\rm cdw}$ to be infinitesimally small but non-zero, or by shifting by infinitesimal amount the momentum $(0,Q)$ from the distance between hot points. Because we analyze the emergence of the static CDW order
and the anti-nesting between regions 3 and 4 is only approximate
, the correct regularization procedure is to shift the momentum.  One can easily make sure that this is equivalent to keeping the integrand as in (\ref{y_3}) and integrating first over $p_x$ and then over $\Omega$.
 Because the integrand vanishes at larger $|p_x|$, the integral over $p_x$ over real axis can be extended in a standard way onto
 a complex plane of $p_x$ and the integration contour can be closed in the half-plane where there is no double pole. The spin-fluctuation propagator
  $\chi$ depends on $p_x$ and has poles in both half-plane of complex $p_x$.  Taking the contributions from the poles in the two $\chi$'s in  the half-plane where there is no double pole, and integrating then over $p_y$ and $\Omega_m$ (in any order), we obtain~\cite{note}
 \begin{align}
\chi_{\rm com} (k_x,k_x',0) = \chi (k_x-k_x', 0)  A(k_x\xi, k_x'\xi, \frac{1 + \lambda}{\lambda}),
\label{ch_3}
\end{align}
The function $A(x,y,z)$ is the scaling function of all three arguments and is $O(1)$ when the arguments are of order one.  When $x=k_x\xi$ and $y=k_x'\xi$ are non-zero, $A(x,y,z)$ evolves but remains close to  $A(0,0,z)$ for
   relevant $ x,y = O(1)$.   As a result, to good accuracy, $\chi_{\rm com} (k_x,k'_x,0)$ and
 $\chi (k_x-k'_x,0)$ differ just by a constant $A (0,0,z)\equiv A(z)$.
 The evaluation of $A(z)$ yields  $A(1) =0.11\bar g/ (\pi \gamma v_F^2) = 0.084$, $A(z \gg1) \approx 1/(2z)$.
 At weak coupling [small $\lambda$ and hence large $z=(1+\lambda)/\lambda$], $A(z)$ is small, but at $\lambda \geq 1$ (hence smaller $z$), $A(z)$ becomes
  of order one. We plot $A(z)$ in Fig.~3.
\begin{figure}
 \includegraphics[width=\columnwidth]{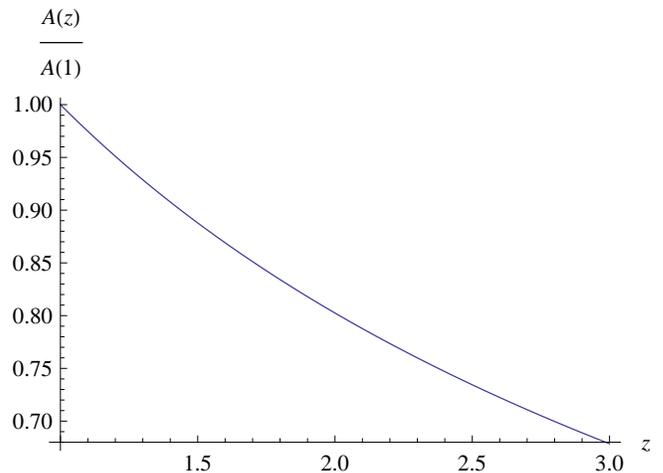}
 \caption{The plot of $A(z)/A(1)=A(0,0,z)/A(0,0,1)$ as a function of $z\equiv (1+\lambda)/\lambda$.}
 \end{figure}

 Note in passing that the value of $A (z)$  can be further increased if we abandon the self-consistent approach, in which $\gamma v^2_F = (4/\pi) {\bar g}$
   and
  assume that
    CDW-x order emerges from some pre-existing pseudogap state which additionally reduces the Landau damping coefficient~\cite{flstar,atkinson,tigran} due to reduction of a
     low-energy fermionic spectral weight in the hot regions.  Because $A(z)$ is inversely proportional to $\gamma$,
      it increases when $\gamma$ gets smaller. This in turn increases $\chi_{\rm com}$ compared with $\chi$.

 To obtain $T_{\rm sc}$ and $T_{\rm cdw}$, one needs to solve Eqs. (\ref{ch_a}). Because $\Phi_{\rm sc} (k)$  has a finite value at $k=0$ and relevant momentum deviations from hot spots are small (of order $\xi^{-1}$), one can safely approximate $\Phi_{\rm sc} (k)$ by $\Phi_{\rm sc} (k_h)$ and explicitly integrate over $k_x-k_x'$ in the spin
  susceptibility.  This leads to a familiar Fermi-liquid result
\beq
 1 = \frac{\lambda}{1+\lambda} \log{\frac{{\bar \omega}_{\rm sf}}{T_{\rm sc}}}
 \eeq
  where ${\bar \omega}_{\rm sf}$ differs from $\omega_{\rm sf}$ by a number.
 Hence
 \beq
 T_{\rm sc} = {\bar \omega}_{\rm sf} ~e^{-\frac{1+\lambda}{\lambda}}
  \label{y_5}
  \eeq
  For CDW-x channel, the evaluation of $T_{\rm cdw}$  requires one to solve the actual integral equation in momentum because
   $\chi_{\rm com}$ depends separately on $k_x$ and $k_x'$. But, like we said, for relevant $k_x \xi \sim k_x' \xi \sim 1$, $A(x,y,z)$ can be well approximated by
    $A (0,0,z) \equiv A(z)$.  Using this approximation, we immediately find that $\Phi_{\rm cdw} (k)$ can be also replaced by its value at $k_h$, and
    $T_{\rm cdw}$ is determined from
   \beq
 1 = A\frac{\lambda}{1+\lambda} \log{\frac{{\bar \omega}_{\rm sf}}{T_{\rm cdw}}}.
 \label{12}
 \eeq
 Hence
 \beq
 T_{\rm cdw} = {\bar \omega}_{\rm sf} ~e^{-\frac{1+\lambda}{A\lambda}}
  \label{y_6}
  \eeq
  At small $\lambda$, $T_{\rm cdw}$ is exponentially suppressed compared to $T_{\rm sc}$, but at $\lambda\geq 1$, $T_{\rm cdw}$ is only numerically but not
   parametrically smaller than $T_{\rm sc}$.

\subsection{SC and CDW-x instabilities at larger $\xi$}

The Fermi liquid consideration is useful for the understanding why $T_{\rm cdw}$ becomes comparable to $T_{\rm sc}$ at $\lambda \geq 1$, but it cannot be extended
 to larger $\xi$ and hence larger $\lambda = 3 {\bar g} \xi/(4 \pi v_F)$ because for these $\xi$ the pairing comes from energies larger than $\omega_{\rm sf}$.
  Specifically, there are two characteristic scales in the problem: $\omega_{\rm sf}$, which is the upper boundary for Fermi-liquid behavior, and
   $\omega_0 = 9 {\bar g}/(16 \pi)$, which is the upper boundary for non-Fermi liquid, quantum-critical behavior with $\Sigma (\omega_m) \approx \sgn (\omega_m) (|\omega_m| \omega_0)^{1/2}$.   At frequencies above $\omega_0$, the self-energy preserves its non-Fermi liquid form but gets smaller than the bare $\omega_m$.
   The ratio of the two energies is  $\omega_{\rm sf} /\omega_0=1/(4\lambda^2)$ (Ref.~\onlinecite{acs}).  At $\lambda \leq 1$, they are comparable, but at large $\lambda$, $\omega_{\rm sf} \ll \omega_0$. In this respect, the upper boundary of the Fermi liquid regime is parametrically smaller  at $\lambda\geq1$ than the highest $\omega_m$ up to which $\Sigma(\omega_m)$ is relevant.
     In the Fermi liquid description,
     the gap equation involves only frequencies $\omega_m<\omega_{\rm sf}$ and hence both
     $T_{\rm sc}$ and $T_{\rm cdw}$ scale with $\omega_{\rm sf}$ and vanish at $\lambda =\xi = \infty$. Meanwhile, the quantum-critical behavior of the system extends to $\omega_0$, which remains finite at $\xi=\infty$. Earlier studies of  d-wave superconductivity found~\cite{acf,acn} that $T_{\rm sc} (\lambda = \infty)$ is in fact finite and is of order
     $\omega_0$.  The issue we consider below is whether $T_{\rm cdw}$ also remains of order $\omega_0$ at infinite $\xi$. We argue that it does and the ratio
     $T_{\rm cdw}/T_{\rm sc}$ is larger than in the Fermi liquid regime.

     Solving the linearized equations for
     $\Phi_{\rm sc}(k_x,\omega_m)$ and
     $\Phi_{\rm cdw}(k_x,\omega_m)$ in the quantum-critical regime is rather involved procedure as these equations become integral equations in
      frequency for $\Phi_{\rm sc}(k_x,\omega_m)$ (Refs.~\onlinecite{acs, ms, wang, wang_el, charge, future}) and in both frequency and momentum for
       $\Phi_{\rm cdw}(k_x,\omega_m)$.
        We refrain from presenting the details of the solution of  the integral equations, but rather focus on proving that (i)
         the mean-field  $T_{\rm cdw}$ is non-zero,
        no matter what the magnitude of the composite interaction is, and has power-law rather than exponential dependence on the interaction strength,
          and (ii) the effective coupling in the CDW channel is smaller numerically
         but not parametrically  than that in the SC channel, hence $T_c$ and $T_{\rm cdw}$ differ by coupling-independent  numerical factor.

    To show that $T_{\rm cdw}$ is non-zero for any interaction strength, we set $T=0$ and consider the equation for CDW-x order parameter as
     eigenvalue/eigenfunction equation. We show that the eigenvalue $E_{\rm cdw}$ is  infinite at $T=0$ for any value of the coupling.
     Because the transition occurs when $E=1$ and $E$ decreases as temperature increases, the very fact that $E_{\rm cdw}$ is infinite at $T=0$ implies that the instability temperature $T_{\rm cdw}$ is finite.

  We first briefly demonstrate how this works for superconductivity.
  The eigenvalue equation for SC order parameter is
  \begin{widetext}
   \begin{align}
    E_{\rm sc} \Phi_{\rm sc}(k_x,\omega_m)=\frac{3\bar g}{8\pi^2 v_F}\int d \omega'_m \frac{1}{|\omega_m'| + \sqrt{|\omega_m'|\omega_0}}
    \int {dk_x'\Phi_{\rm sc}(k_x',\omega_m')}\chi(k_x-k_x',\omega_m-\omega_m')\label{yw1}
     \end{align}
   \end{widetext}
     To obtain this equation we used the non-Fermi-liquid form of the self-energy, and integrated over the transverse momentum $k_y$.
    Because the interaction $\chi$ only depends on momentum transfer $k_x-k_x'$, $\Phi_{\rm sc}(k_x,\omega_m) \equiv \Phi_{\rm sc}(\omega_m)$ is a solution.
     Integrating over $k_x$ in the r.h.s. of  Eq.\ (\ref{yw1}) we obtain~\cite{acf}
\begin{align}
E_{\rm sc}\Phi_{\rm sc}(\omega_m)=\frac{1}{4}\int \frac{d\omega_m'}{\sqrt{|\omega_m'|}}\frac{\Phi_{\rm sc}(\omega_m')}{\sqrt{|\omega_m-\omega_m'|}} \frac{1}{1 + \sqrt{|\omega_m'|/\omega_0}}
\label{yw3}
\end{align}
     One can easily verify that $\Phi_{\rm sc}(\omega_m)=1/\sqrt{|\omega_m|}$ is an eigenfunction, and the corresponding  eigenvalue is infinite:
     \begin{align}
     E_{\rm sc}=\frac{1}{2} \int_0^{\omega_0}\frac{d\omega_m'}{|\omega_m'|} = \infty
     \label{yw10}
          \end{align}
       The logarithmical divergence of $E_{\rm sc}$ at $T=0$ resembles that in the standard BCS theory, but the eigenfunction $\Phi_{\rm cdw}\sim 1/\sqrt{|\omega_m|}$ is different.
         The divergence indicates that at $T=0$ the normal state is unstable towards forming a SC condensate, and there exists a finite
         $T_{\rm sc}\sim \omega_0$ at which SC transition occurs.

 We now apply the same logic to the analysis of $E_{\rm cdw}$ for CDW-x order parameter.
 The ladder equation  for $\Phi_{\rm cdw}(\omega_m)$  in the quantum-critical regime is
     \begin{widetext}
     \begin{align}
    E_{\rm cdw} \Phi_{\rm cdw}(k_x,\omega_m)= \frac{3\bar g}{8\pi^2 v_F}\int d \omega'_m \frac{1}{|\omega_m'| + \sqrt{|\omega_m'|\omega_0}}
    \int {dk_x'\Phi_{\rm cdw}(k_x',\omega_m')}\chi_{\rm com}(k_x,k_x',\omega_m,\omega_m'),\label{yw2}
     \end{align}
     where  the composite interaction in the quantum-critical regime is given by
     \begin{align}
     \chi_{\rm com}(k_x,k_x',\omega_m,\omega_m') =&  -\frac{3\bar g}{8\pi^3}\int \frac{d\Omega_m dp_x dp_y}{(i\sgn (\Omega_m) (|\Omega_m|\omega_0)^{1/2} -v_Fp_x)^2}\nonumber\\
      &\times\frac{1}{(p_x-k_x)^2 + p^2_y + \gamma |\Omega_m-\omega_m|} \frac{1}{(p_x-k_x')^2 + p^2_y + \gamma|\Omega_m-\omega'_m|}.
     \end{align}
         \end{widetext}
We assume and then verify that the eigenfunction of Eq. (\ref{yw2}) takes the form
     \begin{align}
     \Phi_{\rm cdw}(k_x,\omega_m)=\frac{1}{\sqrt{|\omega_m|}}~\varphi\({\tilde k},  \sgn{\omega_m}\)
     \label{yw4}
    \end{align}
    where ${\tilde k} = k_x/\sqrt{\gamma|\omega_m|}$.
   In principle, $\varphi (x,y)$ has both even and odd components in both variables.
    However, substituting $\varphi$ into  (\ref{yw2}) we find after simple algebra that divergent ($\int d \omega_m'/|\omega'_m|$)
     contribution to the r.h.s. of  (\ref{yw2}) comes solely from the even component $\varphi (|\tilde k|)$.
   In explicit form we have
       \begin{align}
    E_{\rm cdw} \varphi(|\tilde k|)=\frac{1}{4\pi}\int_{-\omega_0}^{\omega_0} \frac{d\omega_m'}{|\omega_m'|}\int d\tilde k' \tilde K(\tilde k, \tilde k') \varphi(|\tilde k'|),
    \label{yw7}
    \end{align}
    where
        \begin{align}
    \tilde K(\tilde k, \tilde k')=& -\frac{3}{32\pi^2}\int \frac{d x~ dy~ dz~(x^2 - 9{|z|}/64)}{(x^2 + 9{|z|}/64)^2} \nonumber \\
 &\times \frac{1}{(x-\tilde k)^2 + y^2 + |z|}~\frac{1}{(x-\tilde k')^2 + y^2 + |z-1|}.
 \label{yw6}
    \end{align}
 This is integral equation in ${\tilde k}$ with non-singular momentum dependence in $\varphi(|\tilde k|)$.  We verified that for relevant $\tilde k \leq 1$,
 $\varphi(|\tilde k|)$ can be reasonably well approximated by a constant. Specifically, if we substitute $\varphi(\tilde k') = \varphi$ into the r.h.s.\ of (\ref{yw7}) we find that $f(\tilde k)\equiv \int d\tilde k' K(\tilde k',\tilde k)$ is a slowly varying function of $\tilde k$ (see Fig.\ 4).  Taking $f(0)$ for an estimate, we obtain
     \begin{align}
     E_{\rm cdw} =\frac{C}{2}\int_0^{\omega_0}\frac{d\omega_m'}{|\omega_m'|}.
     \label{yw9}
     \end{align}
     where $C =0.96/\pi =0.31$. We see that $E_{\rm cdw}$ diverges logarithmically at $T=0$, like $E_{\rm sc}$. As the consequence, at $T=0$ the system is
      unstable towards forming a CDW-x condensate, hence $T_{\rm cdw}$ must be finite.
      We note, to avoid misunderstanding, that at this (mean-field) level we consider SC and CDW-x instabilities as independent on each other.

      \begin{figure}
      \includegraphics[width=\columnwidth]{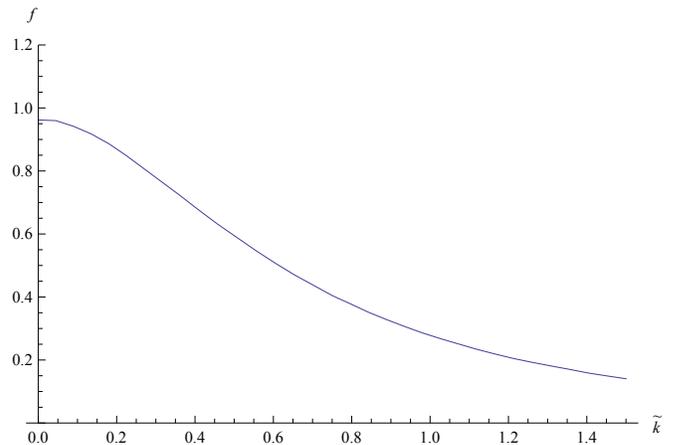}
      \caption{the plot of $f(\tilde k)$ as a function of $\tilde k$. For relevant $\tilde k<1$ it is a slowly-varying function.}
      \end{figure}

Comparing Eq.\ (\ref{yw9}) and Eq.\ (\ref{yw10}), we see that in the quantum-critical regime  the effective dimensionless coupling in the CDW-x channel
 is weaker than that in the SC channel only by
  the numerical factor $C$.  We now use the result of the generic analysis of the SC quantum-critical problem~\cite{future}, which shows that
    the dimensionless coupling $\beta$ appears in the formula for the critical temperature $T_{\rm sc}$, as $\beta^2$ in the overall factor,
     rather than in the exponent.
    Applying this also to $T_{\rm cdw}$, we find that in the quantum-critical regime $T_{\rm cdw}$ is smaller than $T_{\rm sc}$ roughly by $C^2 \sim 0.1$. The ratio  $T_{\rm cdw}/T_{\rm sc}$  can again be enhanced if we assume that CDW-x  emerges from a ``pre-emptive" state in which Landau damping is additionally
     reduced~\cite{flstar,atkinson,tigran}. The ratio $T_{\rm cdw}/T_{\rm sc}$  also get enhanced when we include into the analysis pair-breaking effects by
      thermal fluctuations~\cite{charge}.

 \section{The role of a finite  deviation from nesting/antinsting at hot spots}

 We have demonstrated that for a nested/antinested FS, $T_{\rm cdw}$ is finite for any magnetic correlation length $\xi$, only numerically lower than $T_{\rm sc}$. For a generic FS, there is some small but finite angle between Fermi velocities at, say,  hot spots 1 and 2.

 We note that our analysis for the quantum-critical regime (large $\xi$) did not require any particular nesting condition on the FS -- indeed, at $\xi=\infty$, for a generic FS one can use the same scaling form in Eq.\ (\ref{yw4}) and by the same scaling argument factor out $\int d\omega_m /|\omega_m|$ divergence, only the form of $\tilde K(\tilde k, \tilde k')$ is now more complicated.  Therefore for a generic FS the \mbox{CDW-x} instability still occurs in the quantum-critical regime, although the onset temperature $T_{\rm cdw}$ is smaller than in the perfect nesting/antinesting case.  At a finite $\xi$, the divergence of $E_{\rm cdw}$ at $T=0$ is, however,
  cut, and for any finite deviation from nesting/antinesting limit there exists a finite critical $\xi_{cr}$ at which $T_{\rm cdw}$ vanishes.

  At small deviations from nesting/antinesting critical $\xi_{cr}$ and critical $\lambda_{cr}$ are both small, and the computation of $\xi_{cr}$ can be
   done in the Fermi liquid regime. At small angle $\alpha\equiv v_x/v_y$ between Fermi velocities at hot spots 1 and 2, the fermionic dispersion takes the form $\epsilon_{1,2}({\bf k})=v_F(\pm k_y+\alpha k_x)$
  The convolution of the two Green's functions at hot spots 1 and 2 now gives
 \begin{align}
 T\sum_m^{\omega_{\rm sf}} \int \frac{d\bar k_y}{2\pi} \frac{1}{-(i\omega_m-\alpha \bar k_x)^2+\bar k_y^2}=\log\frac{\omega_{\rm sf}}{\sqrt{T^2+\alpha^2 \bar k_x^2}},
 \end{align}
 where $\bar k_{x,y}\equiv v_Fk_{x,y}/(1+\lambda)$. Taking $\alpha=0$ leads us back to the  $\log (\omega_{\rm sf}/T)$ in Eq.\ (7).
   For $\alpha\neq 0$, the logarithm is cut by ${\bar k}_x$.
  As a consequence, there exists a critical $\xi=\xi_{cr}$ at which $T_{\rm cdw}$ vanishes.
   From Eqs.\ (\ref{y_3}) and (\ref{ch_3}) the typical value  of $k_x$ relevant for CDW-x is $\xi^{-1}$. Then $\xi_{cr}$ is given by
 \begin{align}
   1= A   \frac{\lambda_{cr}}{1+\lambda_{cr}} \log\({\frac{a}{\alpha}\frac{1+\lambda_{cr}}{\lambda_{cr}}}\).
 \end{align}
 where $a = O(1)$ and we recall that $A= A(z)$, $z =(1+\lambda)/\lambda$. Using $A(z \gg 1) \approx 1/(2z)$, we obtain, to logarithmic accuracy,
 \beq
 \lambda_{cr} = \frac{1}{|2 \log {\alpha}|^{1/2}},~~~\xi_{cr} = \frac{4\pi v_F}{3 \bar g} \lambda_{cr}.
 \eeq

Finally, we comment on the validity of the expansion around hot spots, which has been adopted throughout this work.
From Eq.\ (9) we see that the relevant momenta $k_x, k_x'$ (which, we remind, are
deviation from  hot spots 1, 2
along the FS) are of order $\xi^{-1}$.
The typical momenta $p_x$ and $p_y$  in  Eq.\ (8)
are also of order $\xi^{-1}$.
   Taking, e.g., $\xi = 3a$, we obtain that
   typical momentum deviation from a hot spot is $\sim 0.06\times 2\pi/a$. This is a fairly small momentum range.
     As a comparison, for the dispersion taken in e.g., Ref.\ \onlinecite{mike}, the separation between neighboring hot spots
     is a few times higher: $0.2\times 2\pi/a$. In this sense already for $\xi =3a$, SC and CDW instabilities come from the vicinity of hot spots.
     The approximation gets even better when $\xi$ increases.

\section{Summary}

To summarize, we have shown explicitly that
 anti-nesting for a half of hot spots does not prevent the instability towards CDW-x order [the one with momentum $(Q,0)$ or $(0,Q)$]
  as in the strong coupling regime the corresponding  $T_{\rm cdw}$ differs from $T_{\rm sc}$ for d-wave superconductivity only by a constant.
  For the case when the Fermi surface is not perfectly nested/antinested at hot spots,
   we found that at large $\xi$ CDW-x instability still emerges, but terminates at some critical $\xi_{cr} \sim \bar v_F/{\bar g}$.

    The ratio $T_{\rm cdw}/T_{\rm sc}$ is still small numerically -- it is $C^2 \sim 0.1$ at large $\xi$ in fully self-consistent theory
     From this perspective, it is likely that spin-fluctuation exchange is not enough and an additional mechanism,
     i.e., electron-phonon interaction~\cite{ital,mike}, additional softening of fermionic  damping due to pseudogap physics separate from charge order~\cite{flstar,atkinson,tigran}, or Coulomb repulsion between nearest neighbors~\cite{sau} is needed to make CDW-x a strong competitor to d-wave superconductivity in the pseudogap phase of the cuprates.
       Still, from theory perspective, it is essential that the presence of anti-nesting parts on the Fermi surface is not an obstacle for CDW-x instability.

\begin{acknowledgments}
We thank V. Mishra and  M. Norman
 for fruitful discussions. The work was supported by the DOE grant DE-FG02-ER46900.
 \end{acknowledgments}


\begin{thebibliography}{99}
\bibitem{note} The same result can be obtained by integrating over frequency first. For this one has to  split the douple pole by adding an infinitesimal $\epsilon$ to $v_F p_x$ in one of fermionic propagators, combine contributions from the branch cuts in $\chi{'}$s and from the splitted poles in $GG$, and set $\epsilon \to 0$ at the end of the calculation.
\end{thebibliography}
\end{document}